\documentclass[9pt,twocolumn,twoside]{opticajnl}
\journal{opticajournal} 
\pdfoutput=1
\setboolean{shortarticle}{true}

\usepackage{braket}
\usepackage{lineno}
\usepackage{times}
\usepackage{amsmath}
\usepackage{indentfirst}
\usepackage{float}
\usepackage{multirow}
\usepackage[final]{changes}
\usepackage{makecell}
\usepackage{graphicx}
\usepackage{amsthm,amssymb,bbm,newtxmath}
\usepackage{mathrsfs}
\usepackage[noend]{algpseudocode}
\usepackage{array}
\usepackage{booktabs}
\usepackage{stfloats}
\usepackage{makecell}
\usepackage{lineno}

\title{Experimental quantum natural gradient optimization in photonics}

\author[1]{Yizhi Wang}
\author[1]{Shichuan Xue}
\author[1]{Yaxuan Wang}
\author[1]{Jiangfang Ding}
\author[1]{Weixu Shi}
\author[1]{Dongyang Wang}
\author[1]{Yong Liu}
\author[1]{Yingwen Liu}
\author[1]{Xiang Fu}
\author[1]{Guangyao Huang}
\author[1]{Anqi Huang}
\author[1]{Mingtang Deng}
\author[1,*]{Junjie Wu}

\affil[1]{Institute for Quantum Information \& State Key Laboratory of High Performance Computing, College of Computer Science and Technology, National University of Defense Technology, Changsha 410073, China}

\affil[*]{junjiewu@nudt.edu.cn}

\begin{abstract}
Variational quantum algorithms (VQAs) combining the advantages of parameterized quantum circuits and classical optimizers, promise practical quantum applications in the Noisy Intermediate-Scale Quantum era. The performance of VQAs heavily depends on the optimization method. Compared with gradient-free and ordinary gradient descent methods, the quantum natural gradient (QNG), which mirrors the geometric structure of the parameter space, can achieve faster convergence and avoid local minima more easily, thereby reducing the cost of circuit executions. We utilized a fully programmable photonic chip to experimentally estimate the QNG in photonics for the first time.  We obtained the dissociation curve of the He-H$^+$ cation and achieved chemical accuracy, verifying the outperformance of QNG optimization on a photonic device. Our work opens up a vista of utilizing QNG in photonics to implement practical near-term quantum applications.
\end{abstract}

\setboolean{displaycopyright}{true} 

\begin{document}

\maketitle
\textit{Introduction}.—
Recently, a variety of variational quantum algorithms (VQAs) \cite{Cerezo2021} have been proposed with promising applications in quantum chemistry \cite{McArdle2020,Li2019}, materials science \cite{Bauer2020}, quantum machine learning \cite{Zhang2022}, and quantum information processing \cite{Liu2020a,Xue2022a}, etc.
VQAs employ a hybrid quantum-classical framework consisting of three building blocks: 
(\romannumeral1) preparation of parameterized quantum trial state (ansatz),
(\romannumeral2) estimation of the cost function, and 
(\romannumeral3) optimization of quantum circuit parameters utilizing classical optimizers.
Such a framework alleviates the heavy requirement on deep quantum circuits by leveraging additional classical resources,
which promises a leading candidate to bring quantum computing to fruition in practical applications in the Noisy Intermediate-Scale Quantum (NISQ) era \cite{Preskill2018}.

Significant experimental progress has been made in VQA. In photonics, there have been experimental demonstrations ranging from the first variational quantum eigensolver (VQE) \cite{Peruzzo2014} to combined VQEs with phase estimation \cite{Santagati2018} and error mitigation protocols \cite{Lee2022}. However, challenges still remain in classical optimization when scaling up: the optimization can take an extensive number of iterations, yielding a long time and high computational costs until convergence \cite{Bittel2021a}. Prior experimental implementations of VQA in photonics utilized gradient-free optimizers. 
It is known that the gradient information of the cost function can help to guarantee and speed up the convergence of VQAs \cite{Cerezo2021}. 
Traditional vanilla gradient descent methods have improved the VQA performance on various physical platforms, such as superconducting qubits \cite{Kandala2017,Google2020} and trapped ions \cite{Hempel2018,Zhao2022}. 
These methods assume the parameter space to be flat Euclidean, where the steepest descending direction is the negative gradient direction. Physically, however, the parameter distribution is in a Riemannian manifold, and the parametrization is not unique. Different parametrizations vary at a different rate with respect to each parameter, thus distorting distances within the optimization landscape \cite{Amari1998,yamamoto2019natural}. Hence, the distance should be characterized by the KL-divergence rather than Euclidean distance. 
Quantum natural gradient (QNG) \cite{Stokes2020} is such a quantity that uses quantum geometric information with the quantum Fisher information matrix (QFIM), which considers the distance between parameter distributions. QNG optimization moves in the steepest descent direction with respect to the quantum information geometry, corresponding to the real part of the quantum geometric tensor, known as the Fubini-Study metric. QNG has emerged as a superior optimization technique for achieving faster convergence and avoiding local minima \cite{wierichs2020avoiding}.

Using a photonic chip, we experimentally demonstrate an VQE assisted by the QNG-based optimization in photonics for the first time.
The chip has the full ability to (\romannumeral1) prepare an arbitrary quantum quart (ququart) state, (\romannumeral2) perform an arbitrary projective measurement, and (\romannumeral3) estimate the overlap between ququarts.
We compare the convergence performance of vanilla-gradient-based VQE and QNG-based VQE in finding the ground state of the He-H$^+$ cation.
The QNG-based VQE shows superior convergence speed over vanilla gradient descent, achieving more than half of the reduction in optimization iterations.
We employ the Simultaneous Perturbation Stochastic Approximated QNG in VQE to obtain the bond dissociation of the He-H$^+$ cation, and the experimental results are all within chemical accuracy. Our experimental demonstration of QNG-based VQA applications shows that photonic platforms, especially integrated photonics, are competent in making the utmost use of gradient information in the Euclidean and Riemannian manifold space, which could lead to the acceleration of building photonic devices for practical VQA applications.

\begin{figure*}[t]
  \centering\includegraphics{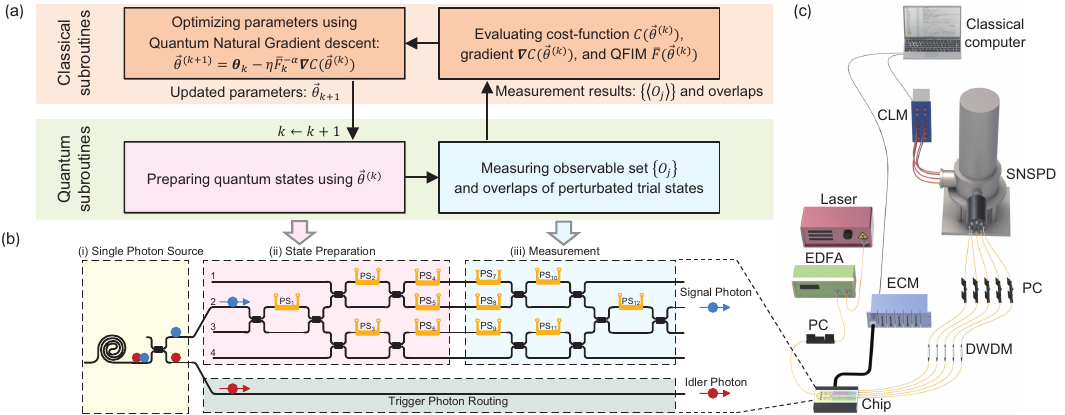}
  \caption{Quantum natural gradient based variational quantum eigensolver and experimental setup. (A) The hybrid quantum-classical workflow of the quantum natural gradient based variational quantum eigensolver. The quantum subroutines employ the chip shown in (B), which mainly includes three functional parts: (\romannumeral1) the heralded single-photon source, (\romannumeral2) preparing arbitrary parameterized ququart states, and (\romannumeral3) performing arbitrary projective measurements.
  (C) Schematic of the chip with the peripheral setup. A tunable continuous wave laser is tuned to a wavelength of 1549.32 nm. The laser is amplified with an optical erbium-doped fiber amplifier (EDFA). The pump laser is launched into the packaged chip through a V-groove fiber array (VGA). Photon pairs generated from the chip are collected from the previous VGA. The signal (1554.13 nm, in blue) and idler (1544.53 nm, in red) photons are separated by dense wavelength-division multiplexing (DWDM) modules. Then photons are detected by superconducting nanowire single photon detectors (SNSPDs), and two-photon coincidence events are recorded by a coincidence logic module (CLM). The polarizations of input and output are tuned by polarization controllers (PCs). An electrical control module (ECM) configures all the on-chip thermo-optic phase shifters. A classical computer compiles optimization algorithms and coordinates control and measurement modules. }
  \label{fig: setup}
\end{figure*}

\textit{Theory of QNG-based VQE}.—
The VQE was first put forward to find the ground state $\ket{\psi_G}$ and ground energy $E_G$ of the Hamiltonian for a given quantum system \cite{Peruzzo2014}. The Hamiltonian is typically given in the form of $H=\sum_j w_j O_j$, which is a weighted sum of Hermitian operators. The Hermitian operators $O_j$ can be measured on the quantum hardware, and their coefficients $w_j$ are stored in classical subroutines.
VQE uses the Rayleigh-Ritz variational principle \cite{Rayleigh1870,Ritz1908} 
\begin{eqnarray}
  E_G \le \braket{\psi(\vec{\theta})\mid H \mid \psi(\vec{\theta})}
\end{eqnarray}
in a hybrid quantum-classical framework, involving iteratively preparing the parameterized quantum state $\ket{\psi(\vec{\theta})}$, estimating the cost function $C(\vec{\theta})=\braket{\psi(\vec{\theta})|H|\psi(\vec{\theta})}$, and optimizing parameters $\vec{\theta}$ on the classical computer to minimize the cost function.
Regarding classical optimization, we mainly considered two types of methods that use gradient information: the vanilla gradient method and the natural gradient method. 
The vanilla gradient method finds a local minimum of $C({\vec{\theta}})$ by moving in the negative gradient direction
\begin{equation}
  \vec{\theta}^{(k+1)} = \vec{\theta}^{(k)}-\eta \boldsymbol{\nabla} C({\vec{\theta}}^{(k)}),
\end{equation}
where $\eta$ is the learning rate and $\boldsymbol{\nabla} C({\vec{\theta}})\in\mathbb{R}$ denotes the gradient of $C({\vec{\theta}})$ with respect to all its parameters.
The parameter vector $\vec{\theta}$ is updated iteratively until convergence.
Nevertheless, such a naive gradient only guarantees the optimal iteration direction under a limited view of the overall cost function landscape. Hence, in physical applications, it tends to fall into local minima.
The natural gradient method is motivated from the perspective of information geometry. It works well for many applications as an alternative to vanilla gradient descent. Mathematically, the natural gradient strategy follows 
\begin{equation}
  \vec{\theta}^{(k+1)}=\vec{\theta}^{(k)}-\eta F(\vec{\theta}^{(k)})^{-1} \boldsymbol{\nabla} C({\vec{\theta}}^{(k)}),
\end{equation}
where $F(\vec{\theta})$ is the quantum Fisher information matrix (QFIM) with the element as 
$F_{ij}(\vec{\theta})=\operatorname{Re}\left\{\left\langle\frac{\partial \psi}{\partial \theta_{i}} \Big|  \frac{\partial \psi}{\partial \theta_{j}}\right\rangle-\left\langle\frac{\partial \psi}{\partial \theta_{i}} \Big| \psi(\vec{\theta})\right\rangle\left\langle\psi(\vec{\theta}) \Big| \frac{\partial \psi}{\partial \theta_{j}}\right\rangle\right\}$.

However, obtaining each of the QFIM elements requires expensive computational costs \cite{wierichs2020avoiding,yamamoto2019natural}. Alternatively, the QFIM can be estimated with the Simultaneous Perturbation Stochastic Approximation (SPSA) algorithm \cite{gacon2021simultaneous}. Instead of computing all elements in each iteration, we can sample the QFIM using two random directions  $\boldsymbol{\Delta}_{1}$  and  $\boldsymbol{\Delta}_{2}$. Given
the Fisher information metric in the state overlap form as
\begin{equation}
  f\left(\vec{\theta}_{1}, \vec{\theta}_{2}\right)=\left|\braket{\psi(\vec{\theta}_1)|{\psi(\vec{\theta}_2)}}\right|^2,
\end{equation}
the estimated QFIM $\tilde{F}(\vec{\theta}^{(k)})$ can be obtained by  $\tilde{F}(\vec{\theta}^{(k)})=\frac{\delta F(\vec{\theta}^{(k)})}{2 \epsilon^{2}} \frac{\boldsymbol{\Delta}_{1} \boldsymbol{\Delta}_{2}^{\text{T}}+\boldsymbol{\Delta}_{2} \boldsymbol{\Delta}_{1}^{\text{T}}}{2}$,
where $\epsilon$ is the perturbation rate, and
\begin{equation}
  \begin{split}
    \delta F(\vec{\theta}^{(k)})&=f\left(\vec{\theta}^{(k)}, \vec{\theta}^{(k)}+\epsilon \boldsymbol{\Delta}_{1}+\epsilon \boldsymbol{\Delta}_{2}\right) 
  -f\left(\vec{\theta}^{(k)}, \vec{\theta}^{(k)}+\epsilon \boldsymbol{\Delta}_{1}\right)\\
  &-f\left(\vec{\theta}^{(k)}, \vec{\theta}^{(k)}-\epsilon \boldsymbol{\Delta}_{1}+\epsilon \boldsymbol{\Delta}_{2}\right) 
  +f\left(\vec{\theta}^{(k)}, \vec{\theta}^{(k)}-\epsilon \boldsymbol{\Delta}_{1}\right). 
  \end{split}
\end{equation}
The point sample $\tilde{F}(\vec{\theta}^{(k)})$ is then combined with all the previous samples in an exponentially smoothed estimator $\bar{F}^{(k)}=\frac{k}{k+1}\bar{F}^{(k-1)}+\frac{1}{k+1}\tilde{F}(\vec{\theta}^{(k)})$.
Then the SPSA-QNG descent update rule is given by
\begin{equation} \label{SPSA-QNG descent update rule}
  \vec{\theta}^{(k+1)}=\vec{\theta}^{(k)}-\eta (\bar{F}^{(k)})^{-\alpha} \boldsymbol{\nabla} C({\vec{\theta}}^{(k)}).
\end{equation}
Here, $\alpha$ is considered a regularization parameter to avoid unstable update caused by possibly ill-conditioned $\tilde{F}(\vec{\theta}^{(k)})$ \cite{Haug2021}. In our experiments, we set $\alpha=0.5$ for a stable half-inversion.

As shown in Fig. \ref{fig: setup}(A), the QNG-based VQE is mainly composed of the following steps \cite{Peruzzo2014,Cerezo2021,Tilly2022}: 

(\romannumeral1) 
(\textit{Quantum subroutine}) Prepare trial states $\ket{\psi(\vec{\theta})}$ using a parameterized circuit $U(\vec{\theta})$. 

(\romannumeral2) 
(\textit{Quantum subroutine}) Estimate groups of observables $\{O_j\}$ that constitute the Hamiltonian $H$, and overlaps of trial states with perturbed parameters as shown in \added{$\delta F(\vec{\theta}^{(k)})$}.

(\romannumeral3) 
(\textit{Classical subroutine}) Evaluate the cost function $C(\vec{\theta})$ and its gradient $\boldsymbol{\nabla} C({\vec{\theta}}^{(k)})$ from the set of measurement results  $\{\braket{O_j}\}$, and evaluate the QFIM $\tilde{F}(\vec{\theta}^{(k)})$ from the overlaps of perturbed trial states.

(\romannumeral4) 
(\textit{Classical subroutine}) Based on the SPSA-QNG descent update rule in Eq. (\ref{SPSA-QNG descent update rule}), update the parameters $\vec{\theta}$ to minimize the expectation value.

(\romannumeral5) Repeat steps (\romannumeral1)-(\romannumeral4) until the desired accuracy is reached. 

\textit{Experimental methods}.—
We experimentally demonstrate the QNG-based VQE to estimate the ground state energy of the He-H$^+$ cation \cite{Peruzzo2014,Lee2022} in a four-dimensional Hilbert space, using parts of a reconfigurable silicon photonic chip \cite{Xue2022} that can high precisely (\romannumeral1) prepare arbitrary ququart states, (\romannumeral2) estimate the overlap between ququarts, and (\romannumeral3) perform an arbitrary projective measurement. The Pauli operators and coefficients constituting the He-H$^+$ Hamiltonian with respect to different interatomic distances are listed in Table S1 of Supplement 1.
The schematic of the chip and the external setup is shown in Fig. \ref{fig: setup}. 
We mainly use three functional parts of the chip. 
In the single-photon source part, a spiral-waveguide spontaneous four-wave mixing (SFWM) photon-pair source \cite{Silverstone2014} is pumped. 
The post-selected signal and idler photons are sent to the state preparing part and passed directly to the detectors, respectively. One idler photon is detected to herald a single signal photon sent to the following operations. The average coincidence rate is $\sim$4.5 kHz.

With the state preparation circuit consisting of three programmable Mach-Zehnder interferometers and three external phase shifters, an arbitrary trial ququart state $\ket{\psi(\vec{\theta})}$ can be generated using the four-dimensional path mode of the heralded photon \cite{Reck1994}. 
Together with the following measurement part, the overlap 
\begin{equation}
  \text{D}(\ket{\psi(\vec{\theta}_1)},\ket{\psi(\vec{\theta}_2)})=\left|{\braket{\psi(\vec{\theta}_1)|\psi(\vec{\theta}_2)}}\right|^2
\end{equation}
between two parameterized ququarts $\ket{\psi(\vec{\theta}_1)}$ and $\ket{\psi(\vec{\theta}_2)}$ can be estimated.
The configuration methods for the on-chip phase shifters (${\rm PS}_{1}\sim{\rm PS}_{12}$) to prepare the trial state $\ket{\psi(\vec{\theta})}$ and estimate the ququart overlap $\text{D}(\ket{\psi(\vec{\theta}_1)},\ket{\psi(\vec{\theta}_2)})$ are detailed in Section 1 of Supplement 1.

We are able to further estimate the result of arbitrary projective measurement.
Given a projective measurement described by a Hermitian operator $O$ and the observable has a spectral decomposition $O = \sum_{j} {m_j P_j}$. Here $P_j = \ket{m_j}\bra{m_j}$ is the projector onto the $j$-th eigenspace of $O$ with eigenvalue $m_j$ and corresponding normalized eigenvector $\ket{m_j}$ \cite{Nielsen2010}. 
We can estimate the overlap between $\ket{\psi(\vec{\theta})}$ and each eigenspace $\text{D}(\ket{\psi(\vec{\theta})},\ket{m_j})$. Then the average value of the measurement of $O$ upon $\ket{\psi(\vec{\theta})}$ can be obtained by 
\begin{equation}
  \text{E}(O)=\braket{\psi(\vec{\theta})\mid O\mid \psi(\vec{\theta})}=\sum\nolimits_j m_j \text{D}(\ket{\psi(\vec{\theta})},\ket{m_j}).
\end{equation}
Accordingly, given the Hamiltonian in the form of $H=\sum_j w_j O_j$, the cost function value $C(\vec{\theta})$ 
can be obtained by weighted summing all the $\text{E}(O_j)$. 

The full abilities of the chip for quantum information processing allow us to extract the analytical gradient in Euclidean space and QFIM from the information geometry. 
The partial derivative of the cost function $C({\vec{\theta}})$ with respect to the photonics phase shift $\theta^{(k)}$ can be estimated using the hardware-friendly parameter-shift rule \cite{Mari2021a}
\begin{equation}
  \frac{\partial C({\vec{\theta}})}{\partial\theta_j}=\frac{C({\vec{\theta}_{j+}})-C({\vec{\theta}_{j-}})}{\sqrt{2}}, 
\end{equation}
with $\vec{\theta}_{j\pm}=\vec{\theta}\pm\frac{\pi}{4}\boldsymbol{e}_j$. Here, $\boldsymbol{e}_j$ is a unit vector with 1 as its $j$-th element and 0 otherwise. Thus, we estimate the analytical gradient $\frac{\partial C({\vec{\theta}})}{\partial\theta_j}$ with two additional measurements that only change the original phase shift $\theta_j$ to $\theta_j+\frac{\pi}{4}$ and $\theta_j-\frac{\pi}{4}$, respectively. Reviewing \added{Eq. (\ref{SPSA-QNG descent update rule})}, we can approximate the QFIM by obtaining the Fisher information metrics from the four overlaps of states with different perturbations.

\begin{figure}[t]
  \centering\includegraphics{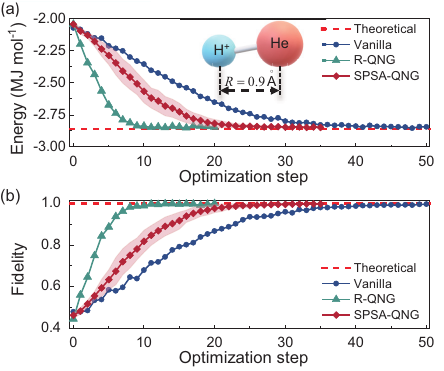}
  \caption{Finding the ground state of the He–H$^+$ cation at the interatomic distance of $R=0.9 \, \mathring{\text{A}}$. A comparison of the convergence performance of the evolution of (A) energy and (B) fidelity using vanilla gradient descent (Vanilla), rigorous quantum natural gradient descent (R-QNG), and Simultaneous Perturbation Stochastic Approximated quantum natural gradient descent (SPSA-QNG). For SPSA-QNG, we repeat the optimization ten times and plot the average and standard deviation (shaded area) of the energy and fidelity in these ten optimizations.} 
  \label{fig: R90}
\end{figure}
\textit{Experimental results}.—
Before running the variational application, we first characterize the high-precision of our chip with one thousand quantum state preparation and measurement experiments.
In each experiment, we program the preparation part of the chip to generate the target random ququart state $\ket{\psi_{\text{the}}}$, and the measurement part to estimate the overlap between the generated ququart $\ket{\psi_{\text{exp}}}$ and $\ket{\psi_{\text{the}}}$.
After collecting the heralded photons from the four output ports, the overlap $\text{D}(\ket{\psi_{\text{the}}},\ket{\psi_{\text{exp}}})$ can be obtained from the probability of heralded photons at the second port.
The statistical quantum state fidelity ${F} = \sqrt{\text{D}(\ket{\psi_{\text{the}}},\ket{\psi_{\text{exp}}})}$ \cite{Nielsen2010} reaches as high as $99.77\pm0.11\%$. The histogram of measured fidelities is shown in Fig. S1 of Supplement 1.

By harnessing our photonic chip to obtain gradient and QNG, we experimentally compare the convergence performance of VQE with three types of gradient-based optimizers, using the Hamiltonian for He-H$^+$ at the interatomic distance $R=0.9 \, \mathring{\text{A}}$ and an initial trial superposition state $\ket{\psi_0}=[0.5,0.5,0.5,0.5]^\text{T}$. As illustrated in Fig. \ref{fig: R90} and Table S2, we present the convergence of the trial state energy and fidelity with the theoretical ground state using a learning rate of $\eta=0.05\pi$ and a perturbation of $\epsilon=0.05\pi$. 
The advantage of QNG is clearly its superior convergence speed compared to vanilla gradient descent. It achieves more than half of the reduction in optimization steps until the energy fluctuation is less than $10^{-2}$ MJ mol$^{-1}$. For SPSA-QNG, even though the convergence speed is slightly worse than rigorous QNG, it almost halves the optimization iterations over vanilla gradient descent. 
The average fidelity of the obtained ground states reaches $99.64\pm0.21\%$. 

\begin{figure}[t]
  \centering\includegraphics{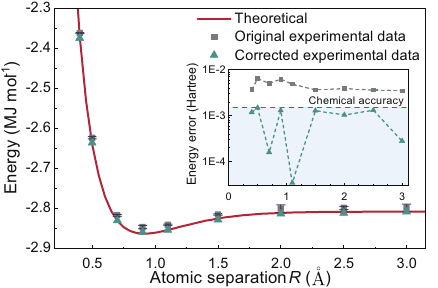}
  \caption{Dissociation curve of the He–H$^+$ cation. For the SPSA-QNG data, the energy and its error bar are obtained by averaging and calculating the standard deviation of results from ten repeated SPSA-QNG VQE runs. In each run, the final energy is obtained by averaging the measured energy values over five steps after convergence. After correcting for a constant systematic error, the data overlaps well with the theoretical energy curve, and the absolute errors (shown in the inset) achieve chemical accuracy.} 
  \label{fig: Curve}
\end{figure}

To obtain the bond dissociation energy of the He-H$^+$ cation, we use the SPSA-QNG-based VQE to search for ground states at a range of nuclear separations $R$. In Fig. \ref{fig: Curve}, the experimental results show good agreement with the theoretical ground-state energies. The experimental data is obtained from the average of ten optimizations at each interatomic distance. After correction for a constant shift $\varepsilon_\text{c}= 0.013$ MJ mol$^{-1}$ \cite{Peruzzo2014}, all absolute errors are less than $3.938$ KJ mol$^{-1}$ ($0.0015$ Hartree), achieving chemical accuracy. For more details, \added{see Section 2}, Table S3, and Fig. S2 of Supplement 1.

\textit{Conclusion}.—
Compared with general gradient-based optimization methods that omit the geometry of the parameter space, quantum natural gradient optimization methods utilize quantum geometric information and adjust the gradient direction accordingly. This approach provides faster convergence and improved robustness, which has been verified in various VQA applications through numerical simulation \cite{Stokes2020,gacon2021simultaneous,Haug2021}.

We used a silicon photonic chip to implement VQEs that are equipped with QNG.
We experimentally obtained the ground energies of the He-H$^+$ cation at a series of interatomic distances and achieved chemical accuracy. From this, we showed that QNG-based optimization has a superior convergence speed over vanilla gradient descent, achieving almost half of the reduction in optimization iterations.
Our work has demonstrated the feasibility and superiority of the QNG-based optimization method in photonics. It has shed light on implementing practical quantum applications in the NISQ era.

\begin{backmatter}
\bmsection{Funding} National Natural Science Foundation of China (62061136011, 62105366, and 62075243).

\deleted{\bmsection{Acknowledgments} The authors acknowledge support from other members of \texttt{QUANTA} (\texttt{QU}antum f\texttt{AN}s from I\texttt{T} \texttt{A}rea) group. }

\bmsection{Disclosures} The authors declare no conflicts of interest.

\bmsection{Data availability} Data underlying the results presented in this paper are not publicly available at this time but may be obtained from the authors upon reasonable request.

\bmsection{Supplemental document}
See Supplement 1 for supporting content. 

\end{backmatter}

\bibliography{ms}


\end{document}


\maketitle

\section{Supplementary method: On-chip preparation of arbitrary ququart states and estimation of overlap between ququarts}

\begin{figure*}[!h]
  \centering\includegraphics{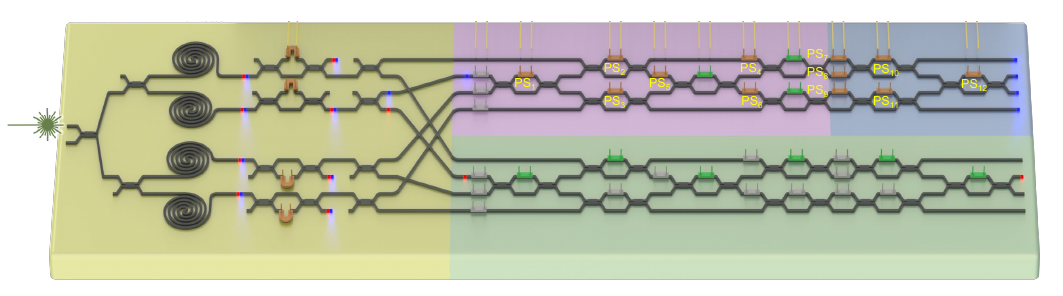}
  \caption{Schematic of the reconfigurable silicon photonic chip used in the experimental quantum natural gradient optimization.  The chip monolithically embeds 4 spontaneous four-wave mixing photon-pair sources, 40 reconfigurable thermo-optic phase shifters, and 44 multimode interferometers. In our experiments, the phase shifters colored in gray stay as spares. We set the phase shifters colored in green as $\pi$. The 12 phase shifters colored in brown are labeled as $\text{PS}_1\sim \text{PS}_{12}$ in the main text Fig. 1, which are used to prepare ququart states and perform projective measurement.}
  \label{fig: chip}
\end{figure*}

We use parts of a reconfigurable silicon photonic chip \cite{Xue2022} to perform experimental demonstrations. In the heralded single-photon source part, four spiral-waveguide spontaneous four-wave mixing photon-pair sources \cite{Silverstone2014} are coherently pumped. By configuring the four asymmetric Mach-Zehnder interferometers \cite{Liu2020}, the photon pairs from the second source are retained while that from the other three sources are filtered out of the chip. The post-selected signal and idler photons are sent to the ququart state preparation part and passed through the trigger routing part to the detectors, respectively.

In the state preparation part, we can prepare an arbitrary ququart state using the heralded photon. Note that any ququart state can be decomposed into the parametric form \cite{Reck1994}: 
\begin{equation} \label{Eq: arbitrary ququart form}
  \begin{split}
    \ket{\psi(\vec{\theta})}
    &=\ket{\psi(\theta_{1},\theta_{2},\theta_{3},\phi_{1},\phi_{2},\phi_{3},\phi_{4})} \\
    &= e^{i\phi_{1}} \sin\theta_{1}\sin\theta_{2} \ket{1} + 
    e^{i\phi_{2}} \sin\theta_{1}\cos\theta_{2} \ket{2} 
    + e^{i\phi_{3}} \cos\theta_{1}\sin\theta_{3} \ket{3} + 
    e^{i\phi_{4}} \cos\theta_{1}\cos\theta_{3} \ket{4},
  \end{split}
\end{equation}
where $\ket{k}$ ($k$=1,2,3,4) is the path mode of the heralded photon.
With the circuit consisting of three programmable Mach-Zehnder interferometers (of which the phase shifters are ${\rm PS}_{1}$, ${\rm PS}_{2}$, and ${\rm PS}_{3}$) and another three extra phase shifters (${\rm PS}_{4}$, ${\rm PS}_{5}$, and ${\rm PS}_{6}$), we can generate the ququart state given in the form of Eq.(\ref{Eq: arbitrary ququart form}) as $\ket{\psi_\text{S}(\vec{\theta}_\text{S})}=\ket{\psi_\text{S}(\theta_\text{S1},\theta_\text{S2},\theta_\text{S3},\phi_\text{S1},\phi_\text{S2},\phi_\text{S3},\phi_\text{S4})}$ by setting the phase shifters as:
\begin{equation}
  \begin{tabular}{ll}
    $\varphi_{{\rm PS}_{1}}=2\theta_\text{S1},$ & $\varphi_{{\rm PS}_{4}}=\phi_\text{S1}-\phi_\text{S4}-\theta_\text{S2}+\theta_\text{S3}+\frac{\pi}{2},$ \\
    $\varphi_{{\rm PS}_{2}}=\pi+2\theta_\text{S2},$ & $\varphi_{{\rm PS}_{5}}=\phi_\text{S2}-\phi_\text{S4}-\theta_\text{S2}+\theta_\text{S3}+\frac{\pi}{2},$ \\
    $\varphi_{{\rm PS}_{3}}=2\theta_\text{S3},$ & $\varphi_{{\rm PS}_{6}}=\phi_\text{S3}-\phi_\text{S4}.$\\
  \end{tabular}
\end{equation}
Then the ququart preparing operation on the heralded photon $U_{\text{prep}}=U_{\text{prep}}(\varphi_{{\rm PS}_{1}},\varphi_{{\rm PS}_{2}},\varphi_{{\rm PS}_{3}},\varphi_{{\rm PS}_{4}},\varphi_{{\rm PS}_{5}},\varphi_{{\rm PS}_{6}})$ 
generates the evolved state $\ket{\psi_{\text{prep}}}=U_{\text{prep}}\ket{2}$, which is equal to $\ket{\psi_\text{S}(\vec{\theta}_\text{S})}$ up to a global phase factor $e^{-i(\theta_\text{S1}+\theta_\text{S3}-\phi_\text{S4}+\pi)}$. 
For example, in our experiments to find ground states of the He-H$^+$ cation, the phase shifts for the initial trial superposition state $\ket{\psi_0}=[0.5,0.5,0.5,0.5]^\text{T}$ are $\vec{\varphi}_0=[\frac{\pi}{2},\frac{3\pi}{2},\frac{\pi}{2},\frac{\pi}{2},\frac{\pi}{2},0]^\text{T}$.

The overlap
\begin{equation}
  \text{D}(\ket{\psi_\text{S}(\vec{\theta}_\text{S})},\ket{\psi_\text{M}(\vec{\theta}_\text{M})})=\left|{\braket{\psi_{\text{\text{S}}}(\vec{\theta}_\text{S})|\psi_\text{M}(\vec{\theta}_\text{M})}}\right|^2
\end{equation}
between the ququarts $\ket{\psi_\text{S}(\vec{\theta}_\text{S})}$ and $\ket{\psi_\text{M}(\vec{\theta}_\text{M})}=\ket{\psi_\text{M}(\theta_\text{M1},\theta_\text{M2},\theta_\text{M3},\phi_\text{M1},\phi_\text{M2},\phi_\text{M3},\phi_\text{M4})}$, can be estimated by using the following measurement circuit to implement the inverse of the preparation operation of $\ket{\psi_\text{M}(\vec{\theta}_\text{M})}$. When the phase shifts for the following six phase shifters are:
\begin{equation}
  \begin{tabular}{ll}
    $\varphi_{{\rm PS}_{7}}=-\phi_\text{M1}+\phi_\text{M4}-\theta_\text{M2}+\theta_\text{M3}+\frac{\pi}{2},$ & $\varphi_{{\rm PS}_{10}}=\pi+2\theta_\text{M2},$ \\
    $\varphi_{{\rm PS}_{8}}=-\phi_\text{M2}+\phi_\text{M4}-\theta_\text{M2}+\theta_\text{M3}+\frac{\pi}{2},$ &  $\varphi_{{\rm PS}_{11}}=2\theta_\text{M3},$ \\
    $\varphi_{{\rm PS}_{9}}=-\phi_\text{M3}+\phi_\text{M4},$ & $\varphi_{{\rm PS}_{12}}=2\theta_\text{M1},$\\
  \end{tabular}
\end{equation}
the operation $U_\text{proj}=U_\text{proj}(\varphi_{{\rm PS}_{7}},\varphi_{{\rm PS}_{8}},\varphi_{{\rm PS}_{9}},\varphi_{{\rm PS}_{10}},\varphi_{{\rm PS}_{11}},\varphi_{{\rm PS}_{12}})$
has the relation $U_\text{proj}\ket{\psi_\text{M}(\vec{\theta}_\text{M})}=e^{i(\theta_\text{M1}+\theta_\text{M3}+\phi_\text{M4}+\pi)} \ket{2}$.
Then the detection probability of heralded photon at the second output port is:
\begin{equation}\label{Eq: inner product}
  \begin{split}
    \text{P}_{(2)} =& \left|{ \bra{2}U_\text{proj}\ket{\psi_{\text{prep}}} }\right|^2 \\
    =&\left|{\bra{\psi_\text{M}(\vec{\theta}_\text{M})}{U_\text{proj}}^\dagger U_\text{proj}\ket{\psi_\text{S}(\vec{\theta}_\text{S})}}\right|^2\\
    =&\left|{\braket{\psi_\text{M}(\vec{\theta}_\text{M})|\psi_\text{S}(\vec{\theta}_\text{S})}}\right|^2\\
    =&\text{D}(\ket{\psi_\text{S}(\vec{\theta}_\text{S})},\ket{\psi_\text{M}(\vec{\theta}_\text{M})}).
  \end{split}
\end{equation}
Accordingly, the overlap between arbitrary ququarts can be experimentally estimated. Note that in experiments, the phase shifts for $\text{PS}_4 \sim \text{PS}_6$ should add $\pi$ to offset the extra phase caused by the three "green" shifters in the state preparation part.

\section{\added{Supplementary method: Estimating the systematic error in measuring Hamiltonians of the He-H$^+$ cation}}

\added{The existence of systematic errors is common when implementing VQEs on NISQ devices, from photonics \cite{Peruzzo2014} and superconducting qubits \cite{Kandala2019} to trapped ions \cite{Zhao2022}. Systematic effects, such as imperfections in the fabrication of the quantum circuit, control noise, and residual crosstalk, contribute to the shift of the measured results of the Hamiltonian. For the specific group of Hamiltonians, systematic errors often manifest as a constant shift, and correction of systematic errors requires first estimating the shift. For example, Ref. \cite{Peruzzo2014} observed a constant and reproducible small shift from the experimentally obtained bond dissociation curve and the theoretical curve. They shifted the experimental results by this constant. Ref. \cite{Zhao2022} shifted all the measured energies so that the energy at a specific interatomic distance matches the simulation energy. In fact, correction methods of Refs. \cite{Peruzzo2014} and \cite{Zhao2022} will mix two sources of error: the systematic error and the optimization (algorithm) error. This is because the VQE optimization results may converge to inaccurate ground states, and the experimentally obtained ground energies will also include the optimization error.}

\added{As shown in the inset of Fig. 3 in the main text, we observed small energy errors within the same order of magnitude in all the experimentally obtained ground energies of the He-H$^+$ cation with respect to different interatomic distances. Systematic effects, such as imperfections in the fabrication of the photonic circuit, control noise of the electrical control module, and crosstalk of the phase shifters, contribute to the shift of the measured results of Pauli strings. The weighted sum of these shifted measurement values of Pauli strings leads to the systematic error of our experimental results.}

\added{To estimate the exact systematic error, we exclude the optimization error by measuring the ground energy using the calculated parameters for the exact ground state (obtained by exact diagonalization). We determine the systematic correction value by measuring the ground energy for the exact ground state of the He-H$^{+}$ cation at R = 0.9 $\mathring{\text{A}}$. The error between the measured value (2.8495 MJ mol$^{-1}$) and theoretical value (2.8626 MJ mol$^{-1}$) is 0.0131 MJ mol$^{-1}$. This measured error gives an estimation of the systematic error at different interatomic distances. After correcting for this constant correction value, our experimental data achieves chemical accuracy.}

\clearpage
\section{Supplementary  Tables}
\begin{table}[ht]
  \centering
  \caption{The table of Pauli strings and their coefficients constituting the He-H$^+$ Hamiltonian with respect to various interatomic distance $R$ \cite{Peruzzo2014}, of which the ground states are experimentally solved in our demonstrations.}
    \begin{tabular}{|c|c|c|c|c|c|c|c|c|c|}
      \hline
      R($\mathring{\text{A}}$)     & II    & IX    & IZ    & XI    & XX    & XZ    & ZI    & ZX    & ZZ \\
      \hline
      0.4   & -1.3119 & -0.1396 & -1.7568 & -0.1396 & 0.3352 & 0.1396 & -1.7568 & 0.1396 & 0.0969 \\
      0.5   & -2.3275 & -0.157 & -1.5236 & -0.157 & 0.3309 & 0.157 & -1.5236 & 0.157 & 0.1115 \\
      0.7   & -3.3893 & -0.1968 & -1.2073 & -0.1968 & 0.3052 & 0.1968 & -1.2073 & 0.1968 & 0.1626 \\
      0.9   & -3.8505 & -0.2288 & -1.0466 & -0.2288 & 0.2613 & 0.2288 & -1.0466 & 0.2288 & 0.2356 \\
      1.1   & -4.0539 & -0.243 & -0.982 & -0.243 & 0.2053 & 0.243 & -0.982 & 0.243 & 0.3225 \\
      1.5   & -4.1594 & -0.2086 & -0.991 & -0.2086 & 0.0948 & 0.2086 & -0.991 & 0.2086 & 0.4945 \\
      2     & -4.1347 & -0.1119 & -1.0605 & -0.1119 & 0.0212 & 0.1119 & -1.0605 & 0.1119 & 0.6342 \\
      2.5   & -4.0918 & -0.0454 & -1.1128 & -0.0454 & 0.0032 & 0.0454 & -1.1128 & 0.0454 & 0.701 \\
      3     & -4.0578 & -0.0159 & -1.1482 & -0.0159 & 0.0004 & 0.0159 & -1.1482 & 0.0159 & 0.7385 \\
      \hline
    \end{tabular}%
  \label{tab:Hamiltonian}%
\end{table}%

\begin{table}[hb]
  \centering
  \caption{Experimental data for the vanilla gradient descent, rigorous quantum natural gradient descent (R-QNG), and SPSA quantum natural gradient descent (SPSA-QNG) optimizations in finding ground states of the He-H$^+$ cation at interatomic distance \replaced{$R=0.9\,\mathring{\text{A}}$}{$R=90\,\mathring{\text{A}}$} shown in Fig. 2. The unit of energy is MJ mol$^{-1}$.} 
  \scalebox{0.45}{
    \begin{tabular}{|c|c|c|cc|cccccccccccccccccccccc|}
    \hline
    \multirow{2}[4]{*}{Steps} & \multicolumn{2}{c|}{Vanilla} & \multicolumn{2}{c|}{R-QNG} & \multicolumn{2}{c|}{SPSA-QNG 1} & \multicolumn{2}{c|}{SPSA-QNG 2} & \multicolumn{2}{c|}{SPSA-QNG 3} & \multicolumn{2}{c|}{SPSA-QNG 4} & \multicolumn{2}{c|}{SPSA-QNG 5} & \multicolumn{2}{c|}{SPSA-QNG 6} & \multicolumn{2}{c|}{SPSA-QNG 7} & \multicolumn{2}{c|}{SPSA-QNG 8} & \multicolumn{2}{c|}{SPSA-QNG 9} & \multicolumn{2}{c|}{SPSA-QNG 10} & \multicolumn{2}{c|}{SPSA-QNG Average} \\
\cline{2-27}          & Energy & Fidelity & \multicolumn{1}{c|}{Energy} & Fidelity & \multicolumn{1}{c|}{Energy} & \multicolumn{1}{c|}{Fidelity} & \multicolumn{1}{c|}{Energy} & \multicolumn{1}{c|}{Fidelity} & \multicolumn{1}{c|}{Energy} & \multicolumn{1}{c|}{Fidelity} & \multicolumn{1}{c|}{Energy} & \multicolumn{1}{c|}{Fidelity} & \multicolumn{1}{c|}{Energy} & \multicolumn{1}{c|}{Fidelity} & \multicolumn{1}{c|}{Energy} & \multicolumn{1}{c|}{Fidelity} & \multicolumn{1}{c|}{Energy} & \multicolumn{1}{c|}{Fidelity} & \multicolumn{1}{c|}{Energy} & \multicolumn{1}{c|}{Fidelity} & \multicolumn{1}{c|}{Energy} & \multicolumn{1}{c|}{Fidelity} & \multicolumn{1}{c|}{Energy} & \multicolumn{1}{c|}{Fidelity} & \multicolumn{1}{c|}{Energy} & Fidelity \\
    \hline
    0     & -2.072  & 47.86\% & \multicolumn{1}{c|}{-2.042 } & 44.38\% & \multicolumn{1}{c|}{-2.043 } & \multicolumn{1}{c|}{45.53\%} & \multicolumn{1}{c|}{-2.056 } & \multicolumn{1}{c|}{47.16\%} & \multicolumn{1}{c|}{-2.025 } & \multicolumn{1}{c|}{44.50\%} & \multicolumn{1}{c|}{-2.054 } & \multicolumn{1}{c|}{46.61\%} & \multicolumn{1}{c|}{-2.046 } & \multicolumn{1}{c|}{46.24\%} & \multicolumn{1}{c|}{-2.051 } & \multicolumn{1}{c|}{46.11\%} & \multicolumn{1}{c|}{-2.034 } & \multicolumn{1}{c|}{48.52\%} & \multicolumn{1}{c|}{-2.055 } & \multicolumn{1}{c|}{48.35\%} & \multicolumn{1}{c|}{-2.003 } & \multicolumn{1}{c|}{45.02\%} & \multicolumn{1}{c|}{-2.067 } & \multicolumn{1}{c|}{44.62\%} & \multicolumn{1}{c|}{-2.043$\pm$0.018 } & 46.27$\pm$1.35\%  \\
    \hline
    1     & -2.083  & 47.95\% & \multicolumn{1}{c|}{-2.198 } & 55.92\% & \multicolumn{1}{c|}{-2.109 } & \multicolumn{1}{c|}{48.32\%} & \multicolumn{1}{c|}{-2.067 } & \multicolumn{1}{c|}{47.95\%} & \multicolumn{1}{c|}{-2.064 } & \multicolumn{1}{c|}{49.61\%} & \multicolumn{1}{c|}{-2.097 } & \multicolumn{1}{c|}{47.64\%} & \multicolumn{1}{c|}{-2.071 } & \multicolumn{1}{c|}{48.04\%} & \multicolumn{1}{c|}{-2.116 } & \multicolumn{1}{c|}{45.59\%} & \multicolumn{1}{c|}{-2.126 } & \multicolumn{1}{c|}{47.52\%} & \multicolumn{1}{c|}{-2.080 } & \multicolumn{1}{c|}{46.74\%} & \multicolumn{1}{c|}{-2.093 } & \multicolumn{1}{c|}{51.16\%} & \multicolumn{1}{c|}{-2.134 } & \multicolumn{1}{c|}{49.97\%} & \multicolumn{1}{c|}{-2.096$\pm$0.024 } & 48.25$\pm$1.54\%  \\
    \hline
    2     & -2.119  & 48.50\% & \multicolumn{1}{c|}{-2.344 } & 64.53\% & \multicolumn{1}{c|}{-2.138 } & \multicolumn{1}{c|}{51.08\%} & \multicolumn{1}{c|}{-2.118 } & \multicolumn{1}{c|}{50.05\%} & \multicolumn{1}{c|}{-2.081 } & \multicolumn{1}{c|}{50.10\%} & \multicolumn{1}{c|}{-2.140 } & \multicolumn{1}{c|}{51.75\%} & \multicolumn{1}{c|}{-2.119 } & \multicolumn{1}{c|}{49.87\%} & \multicolumn{1}{c|}{-2.128 } & \multicolumn{1}{c|}{52.68\%} & \multicolumn{1}{c|}{-2.170 } & \multicolumn{1}{c|}{56.28\%} & \multicolumn{1}{c|}{-2.146 } & \multicolumn{1}{c|}{51.48\%} & \multicolumn{1}{c|}{-2.137 } & \multicolumn{1}{c|}{52.36\%} & \multicolumn{1}{c|}{-2.114 } & \multicolumn{1}{c|}{51.98\%} & \multicolumn{1}{c|}{-2.129$\pm$0.022 } & 51.76$\pm$1.77\%  \\
    \hline
    3     & -2.149  & 53.42\% & \multicolumn{1}{c|}{-2.436 } & 74.77\% & \multicolumn{1}{c|}{-2.173 } & \multicolumn{1}{c|}{55.15\%} & \multicolumn{1}{c|}{-2.174 } & \multicolumn{1}{c|}{55.12\%} & \multicolumn{1}{c|}{-2.124 } & \multicolumn{1}{c|}{52.33\%} & \multicolumn{1}{c|}{-2.171 } & \multicolumn{1}{c|}{53.85\%} & \multicolumn{1}{c|}{-2.147 } & \multicolumn{1}{c|}{54.85\%} & \multicolumn{1}{c|}{-2.194 } & \multicolumn{1}{c|}{54.84\%} & \multicolumn{1}{c|}{-2.271 } & \multicolumn{1}{c|}{57.47\%} & \multicolumn{1}{c|}{-2.203 } & \multicolumn{1}{c|}{58.78\%} & \multicolumn{1}{c|}{-2.182 } & \multicolumn{1}{c|}{59.24\%} & \multicolumn{1}{c|}{-2.129 } & \multicolumn{1}{c|}{53.97\%} & \multicolumn{1}{c|}{-2.177$\pm$0.040 } & 55.56$\pm$2.12\%  \\
    \hline
    4     & -2.151  & 53.78\% & \multicolumn{1}{c|}{-2.564 } & 84.31\% & \multicolumn{1}{c|}{-2.234 } & \multicolumn{1}{c|}{60.23\%} & \multicolumn{1}{c|}{-2.214 } & \multicolumn{1}{c|}{58.10\%} & \multicolumn{1}{c|}{-2.135 } & \multicolumn{1}{c|}{54.08\%} & \multicolumn{1}{c|}{-2.248 } & \multicolumn{1}{c|}{59.10\%} & \multicolumn{1}{c|}{-2.198 } & \multicolumn{1}{c|}{54.48\%} & \multicolumn{1}{c|}{-2.261 } & \multicolumn{1}{c|}{60.32\%} & \multicolumn{1}{c|}{-2.343 } & \multicolumn{1}{c|}{64.65\%} & \multicolumn{1}{c|}{-2.287 } & \multicolumn{1}{c|}{66.79\%} & \multicolumn{1}{c|}{-2.259 } & \multicolumn{1}{c|}{61.79\%} & \multicolumn{1}{c|}{-2.173 } & \multicolumn{1}{c|}{54.47\%} & \multicolumn{1}{c|}{-2.235$\pm$0.056 } & 59.40$\pm$4.10\%  \\
    \hline
    5     & -2.213  & 58.48\% & \multicolumn{1}{c|}{-2.674 } & 88.56\% & \multicolumn{1}{c|}{-2.278 } & \multicolumn{1}{c|}{62.22\%} & \multicolumn{1}{c|}{-2.287 } & \multicolumn{1}{c|}{59.24\%} & \multicolumn{1}{c|}{-2.158 } & \multicolumn{1}{c|}{55.02\%} & \multicolumn{1}{c|}{-2.281 } & \multicolumn{1}{c|}{62.36\%} & \multicolumn{1}{c|}{-2.220 } & \multicolumn{1}{c|}{61.86\%} & \multicolumn{1}{c|}{-2.391 } & \multicolumn{1}{c|}{68.73\%} & \multicolumn{1}{c|}{-2.360 } & \multicolumn{1}{c|}{67.62\%} & \multicolumn{1}{c|}{-2.347 } & \multicolumn{1}{c|}{68.39\%} & \multicolumn{1}{c|}{-2.345 } & \multicolumn{1}{c|}{69.69\%} & \multicolumn{1}{c|}{-2.198 } & \multicolumn{1}{c|}{58.40\%} & \multicolumn{1}{c|}{-2.287$\pm$0.072 } & 63.35$\pm$4.78\%  \\
    \hline
    6     & -2.231  & 57.93\% & \multicolumn{1}{c|}{-2.740 } & 93.60\% & \multicolumn{1}{c|}{-2.336 } & \multicolumn{1}{c|}{68.39\%} & \multicolumn{1}{c|}{-2.342 } & \multicolumn{1}{c|}{65.82\%} & \multicolumn{1}{c|}{-2.186 } & \multicolumn{1}{c|}{56.57\%} & \multicolumn{1}{c|}{-2.339 } & \multicolumn{1}{c|}{64.56\%} & \multicolumn{1}{c|}{-2.278 } & \multicolumn{1}{c|}{61.58\%} & \multicolumn{1}{c|}{-2.511 } & \multicolumn{1}{c|}{78.97\%} & \multicolumn{1}{c|}{-2.438 } & \multicolumn{1}{c|}{73.51\%} & \multicolumn{1}{c|}{-2.407 } & \multicolumn{1}{c|}{73.47\%} & \multicolumn{1}{c|}{-2.381 } & \multicolumn{1}{c|}{70.64\%} & \multicolumn{1}{c|}{-2.222 } & \multicolumn{1}{c|}{60.82\%} & \multicolumn{1}{c|}{-2.344$\pm$0.093 } & 67.43$\pm$6.52\%  \\
    \hline
    7     & -2.250  & 59.62\% & \multicolumn{1}{c|}{-2.790 } & 96.07\% & \multicolumn{1}{c|}{-2.470 } & \multicolumn{1}{c|}{77.45\%} & \multicolumn{1}{c|}{-2.407 } & \multicolumn{1}{c|}{67.52\%} & \multicolumn{1}{c|}{-2.239 } & \multicolumn{1}{c|}{62.42\%} & \multicolumn{1}{c|}{-2.367 } & \multicolumn{1}{c|}{69.24\%} & \multicolumn{1}{c|}{-2.310 } & \multicolumn{1}{c|}{65.45\%} & \multicolumn{1}{c|}{-2.576 } & \multicolumn{1}{c|}{83.35\%} & \multicolumn{1}{c|}{-2.508 } & \multicolumn{1}{c|}{78.57\%} & \multicolumn{1}{c|}{-2.450 } & \multicolumn{1}{c|}{75.62\%} & \multicolumn{1}{c|}{-2.412 } & \multicolumn{1}{c|}{71.05\%} & \multicolumn{1}{c|}{-2.350 } & \multicolumn{1}{c|}{68.53\%} & \multicolumn{1}{c|}{-2.409$\pm$0.093 } & 71.92$\pm$6.24\%  \\
    \hline
    8     & -2.299  & 64.48\% & \multicolumn{1}{c|}{-2.813 } & 98.14\% & \multicolumn{1}{c|}{-2.497 } & \multicolumn{1}{c|}{78.73\%} & \multicolumn{1}{c|}{-2.467 } & \multicolumn{1}{c|}{75.32\%} & \multicolumn{1}{c|}{-2.310 } & \multicolumn{1}{c|}{65.96\%} & \multicolumn{1}{c|}{-2.412 } & \multicolumn{1}{c|}{70.05\%} & \multicolumn{1}{c|}{-2.356 } & \multicolumn{1}{c|}{68.81\%} & \multicolumn{1}{c|}{-2.641 } & \multicolumn{1}{c|}{86.25\%} & \multicolumn{1}{c|}{-2.587 } & \multicolumn{1}{c|}{82.75\%} & \multicolumn{1}{c|}{-2.465 } & \multicolumn{1}{c|}{78.00\%} & \multicolumn{1}{c|}{-2.429 } & \multicolumn{1}{c|}{73.74\%} & \multicolumn{1}{c|}{-2.386 } & \multicolumn{1}{c|}{71.09\%} & \multicolumn{1}{c|}{-2.455$\pm$0.096 } & 75.07$\pm$6.09\%  \\
    \hline
    9     & -2.331  & 63.78\% & \multicolumn{1}{c|}{-2.836 } & 98.58\% & \multicolumn{1}{c|}{-2.538 } & \multicolumn{1}{c|}{79.79\%} & \multicolumn{1}{c|}{-2.527 } & \multicolumn{1}{c|}{77.70\%} & \multicolumn{1}{c|}{-2.412 } & \multicolumn{1}{c|}{73.64\%} & \multicolumn{1}{c|}{-2.445 } & \multicolumn{1}{c|}{73.22\%} & \multicolumn{1}{c|}{-2.393 } & \multicolumn{1}{c|}{72.59\%} & \multicolumn{1}{c|}{-2.645 } & \multicolumn{1}{c|}{86.72\%} & \multicolumn{1}{c|}{-2.651 } & \multicolumn{1}{c|}{86.55\%} & \multicolumn{1}{c|}{-2.522 } & \multicolumn{1}{c|}{81.39\%} & \multicolumn{1}{c|}{-2.498 } & \multicolumn{1}{c|}{76.84\%} & \multicolumn{1}{c|}{-2.441 } & \multicolumn{1}{c|}{72.12\%} & \multicolumn{1}{c|}{-2.507$\pm$0.085 } & 78.06$\pm$5.21\%  \\
    \hline
    10    & -2.354  & 67.84\% & \multicolumn{1}{c|}{-2.839 } & 98.71\% & \multicolumn{1}{c|}{-2.590 } & \multicolumn{1}{c|}{86.14\%} & \multicolumn{1}{c|}{-2.587 } & \multicolumn{1}{c|}{82.13\%} & \multicolumn{1}{c|}{-2.535 } & \multicolumn{1}{c|}{80.88\%} & \multicolumn{1}{c|}{-2.470 } & \multicolumn{1}{c|}{75.24\%} & \multicolumn{1}{c|}{-2.421 } & \multicolumn{1}{c|}{72.82\%} & \multicolumn{1}{c|}{-2.703 } & \multicolumn{1}{c|}{88.87\%} & \multicolumn{1}{c|}{-2.707 } & \multicolumn{1}{c|}{89.25\%} & \multicolumn{1}{c|}{-2.613 } & \multicolumn{1}{c|}{83.74\%} & \multicolumn{1}{c|}{-2.554 } & \multicolumn{1}{c|}{81.01\%} & \multicolumn{1}{c|}{-2.489 } & \multicolumn{1}{c|}{77.22\%} & \multicolumn{1}{c|}{-2.567$\pm$0.089 } & 81.73$\pm$5.23\%  \\
    \hline
    11    & -2.402  & 71.35\% & \multicolumn{1}{c|}{-2.849 } & 99.59\% & \multicolumn{1}{c|}{-2.643 } & \multicolumn{1}{c|}{88.16\%} & \multicolumn{1}{c|}{-2.639 } & \multicolumn{1}{c|}{85.14\%} & \multicolumn{1}{c|}{-2.638 } & \multicolumn{1}{c|}{85.62\%} & \multicolumn{1}{c|}{-2.529 } & \multicolumn{1}{c|}{78.16\%} & \multicolumn{1}{c|}{-2.482 } & \multicolumn{1}{c|}{77.19\%} & \multicolumn{1}{c|}{-2.716 } & \multicolumn{1}{c|}{89.65\%} & \multicolumn{1}{c|}{-2.730 } & \multicolumn{1}{c|}{90.70\%} & \multicolumn{1}{c|}{-2.667 } & \multicolumn{1}{c|}{88.49\%} & \multicolumn{1}{c|}{-2.610 } & \multicolumn{1}{c|}{84.04\%} & \multicolumn{1}{c|}{-2.532 } & \multicolumn{1}{c|}{78.55\%} & \multicolumn{1}{c|}{-2.619$\pm$0.077 } & 84.57$\pm$4.75\%  \\
    \hline
    12    & -2.425  & 71.80\% & \multicolumn{1}{c|}{-2.844 } & 99.36\% & \multicolumn{1}{c|}{-2.684 } & \multicolumn{1}{c|}{89.18\%} & \multicolumn{1}{c|}{-2.681 } & \multicolumn{1}{c|}{86.98\%} & \multicolumn{1}{c|}{-2.677 } & \multicolumn{1}{c|}{89.73\%} & \multicolumn{1}{c|}{-2.533 } & \multicolumn{1}{c|}{79.08\%} & \multicolumn{1}{c|}{-2.512 } & \multicolumn{1}{c|}{80.36\%} & \multicolumn{1}{c|}{-2.721 } & \multicolumn{1}{c|}{89.32\%} & \multicolumn{1}{c|}{-2.767 } & \multicolumn{1}{c|}{93.04\%} & \multicolumn{1}{c|}{-2.701 } & \multicolumn{1}{c|}{92.22\%} & \multicolumn{1}{c|}{-2.597 } & \multicolumn{1}{c|}{85.15\%} & \multicolumn{1}{c|}{-2.551 } & \multicolumn{1}{c|}{82.79\%} & \multicolumn{1}{c|}{-2.642$\pm$0.083 } & 86.79$\pm$4.57\%  \\
    \hline
    13    & -2.453  & 74.27\% & \multicolumn{1}{c|}{-2.849 } & 99.75\% & \multicolumn{1}{c|}{-2.705 } & \multicolumn{1}{c|}{90.26\%} & \multicolumn{1}{c|}{-2.716 } & \multicolumn{1}{c|}{91.01\%} & \multicolumn{1}{c|}{-2.733 } & \multicolumn{1}{c|}{91.41\%} & \multicolumn{1}{c|}{-2.593 } & \multicolumn{1}{c|}{82.32\%} & \multicolumn{1}{c|}{-2.563 } & \multicolumn{1}{c|}{83.18\%} & \multicolumn{1}{c|}{-2.749 } & \multicolumn{1}{c|}{92.70\%} & \multicolumn{1}{c|}{-2.785 } & \multicolumn{1}{c|}{95.00\%} & \multicolumn{1}{c|}{-2.747 } & \multicolumn{1}{c|}{93.80\%} & \multicolumn{1}{c|}{-2.635 } & \multicolumn{1}{c|}{85.02\%} & \multicolumn{1}{c|}{-2.594 } & \multicolumn{1}{c|}{84.43\%} & \multicolumn{1}{c|}{-2.682$\pm$0.074 } & 88.91$\pm$4.47\%  \\
    \hline
    14    & -2.488  & 77.62\% & \multicolumn{1}{c|}{-2.840 } & 99.72\% & \multicolumn{1}{c|}{-2.727 } & \multicolumn{1}{c|}{91.45\%} & \multicolumn{1}{c|}{-2.734 } & \multicolumn{1}{c|}{92.34\%} & \multicolumn{1}{c|}{-2.746 } & \multicolumn{1}{c|}{93.98\%} & \multicolumn{1}{c|}{-2.612 } & \multicolumn{1}{c|}{83.70\%} & \multicolumn{1}{c|}{-2.617 } & \multicolumn{1}{c|}{85.44\%} & \multicolumn{1}{c|}{-2.745 } & \multicolumn{1}{c|}{92.69\%} & \multicolumn{1}{c|}{-2.798 } & \multicolumn{1}{c|}{95.51\%} & \multicolumn{1}{c|}{-2.772 } & \multicolumn{1}{c|}{95.32\%} & \multicolumn{1}{c|}{-2.617 } & \multicolumn{1}{c|}{86.44\%} & \multicolumn{1}{c|}{-2.621 } & \multicolumn{1}{c|}{85.42\%} & \multicolumn{1}{c|}{-2.699$\pm$0.070 } & 90.23$\pm$4.28\%  \\
    \hline
    15    & -2.521  & 79.09\% & \multicolumn{1}{c|}{-2.844 } & 99.49\% & \multicolumn{1}{c|}{-2.744 } & \multicolumn{1}{c|}{92.92\%} & \multicolumn{1}{c|}{-2.772 } & \multicolumn{1}{c|}{93.96\%} & \multicolumn{1}{c|}{-2.775 } & \multicolumn{1}{c|}{95.34\%} & \multicolumn{1}{c|}{-2.658 } & \multicolumn{1}{c|}{84.53\%} & \multicolumn{1}{c|}{-2.647 } & \multicolumn{1}{c|}{87.32\%} & \multicolumn{1}{c|}{-2.763 } & \multicolumn{1}{c|}{93.38\%} & \multicolumn{1}{c|}{-2.815 } & \multicolumn{1}{c|}{95.83\%} & \multicolumn{1}{c|}{-2.798 } & \multicolumn{1}{c|}{96.93\%} & \multicolumn{1}{c|}{-2.632 } & \multicolumn{1}{c|}{87.25\%} & \multicolumn{1}{c|}{-2.655 } & \multicolumn{1}{c|}{89.01\%} & \multicolumn{1}{c|}{-2.726$\pm$0.067 } & 91.65$\pm$4.06\%  \\
    \hline
    16    & -2.538  & 80.04\% & \multicolumn{1}{c|}{-2.853 } & 99.65\% & \multicolumn{1}{c|}{-2.772 } & \multicolumn{1}{c|}{94.60\%} & \multicolumn{1}{c|}{-2.790 } & \multicolumn{1}{c|}{95.02\%} & \multicolumn{1}{c|}{-2.806 } & \multicolumn{1}{c|}{95.63\%} & \multicolumn{1}{c|}{-2.677 } & \multicolumn{1}{c|}{88.62\%} & \multicolumn{1}{c|}{-2.669 } & \multicolumn{1}{c|}{89.29\%} & \multicolumn{1}{c|}{-2.790 } & \multicolumn{1}{c|}{94.20\%} & \multicolumn{1}{c|}{-2.813 } & \multicolumn{1}{c|}{97.36\%} & \multicolumn{1}{c|}{-2.810 } & \multicolumn{1}{c|}{96.89\%} & \multicolumn{1}{c|}{-2.756 } & \multicolumn{1}{c|}{95.62\%} & \multicolumn{1}{c|}{-2.705 } & \multicolumn{1}{c|}{90.89\%} & \multicolumn{1}{c|}{-2.759$\pm$0.053 } & 93.81$\pm$2.95\%  \\
    \hline
    17    & -2.565  & 81.03\% & \multicolumn{1}{c|}{-2.837 } & 99.54\% & \multicolumn{1}{c|}{-2.786 } & \multicolumn{1}{c|}{95.54\%} & \multicolumn{1}{c|}{-2.801 } & \multicolumn{1}{c|}{95.51\%} & \multicolumn{1}{c|}{-2.817 } & \multicolumn{1}{c|}{97.69\%} & \multicolumn{1}{c|}{-2.695 } & \multicolumn{1}{c|}{90.32\%} & \multicolumn{1}{c|}{-2.722 } & \multicolumn{1}{c|}{90.33\%} & \multicolumn{1}{c|}{-2.788 } & \multicolumn{1}{c|}{94.73\%} & \multicolumn{1}{c|}{-2.826 } & \multicolumn{1}{c|}{97.99\%} & \multicolumn{1}{c|}{-2.809 } & \multicolumn{1}{c|}{97.71\%} & \multicolumn{1}{c|}{-2.826 } & \multicolumn{1}{c|}{99.04\%} & \multicolumn{1}{c|}{-2.726 } & \multicolumn{1}{c|}{92.56\%} & \multicolumn{1}{c|}{-2.780$\pm$0.046 } & 95.14$\pm$3.00\%  \\
    \hline
    18    & -2.607  & 83.83\% & \multicolumn{1}{c|}{-2.839 } & 98.94\% & \multicolumn{1}{c|}{-2.818 } & \multicolumn{1}{c|}{96.25\%} & \multicolumn{1}{c|}{-2.804 } & \multicolumn{1}{c|}{95.26\%} & \multicolumn{1}{c|}{-2.823 } & \multicolumn{1}{c|}{98.42\%} & \multicolumn{1}{c|}{-2.715 } & \multicolumn{1}{c|}{91.77\%} & \multicolumn{1}{c|}{-2.749 } & \multicolumn{1}{c|}{93.94\%} & \multicolumn{1}{c|}{-2.804 } & \multicolumn{1}{c|}{95.97\%} & \multicolumn{1}{c|}{-2.834 } & \multicolumn{1}{c|}{98.18\%} & \multicolumn{1}{c|}{-2.826 } & \multicolumn{1}{c|}{98.81\%} & \multicolumn{1}{c|}{-2.836 } & \multicolumn{1}{c|}{99.38\%} & \multicolumn{1}{c|}{-2.731 } & \multicolumn{1}{c|}{93.30\%} & \multicolumn{1}{c|}{-2.794$\pm$0.043 } & 96.13$\pm$2.45\%  \\
    \hline
    19    & -2.622  & 85.20\% & \multicolumn{1}{c|}{-2.843 } & 99.50\% & \multicolumn{1}{c|}{-2.822 } & \multicolumn{1}{c|}{98.46\%} & \multicolumn{1}{c|}{-2.828 } & \multicolumn{1}{c|}{96.39\%} & \multicolumn{1}{c|}{-2.830 } & \multicolumn{1}{c|}{98.78\%} & \multicolumn{1}{c|}{-2.739 } & \multicolumn{1}{c|}{92.86\%} & \multicolumn{1}{c|}{-2.765 } & \multicolumn{1}{c|}{93.73\%} & \multicolumn{1}{c|}{-2.814 } & \multicolumn{1}{c|}{95.86\%} & \multicolumn{1}{c|}{-2.842 } & \multicolumn{1}{c|}{98.18\%} & \multicolumn{1}{c|}{-2.830 } & \multicolumn{1}{c|}{98.93\%} & \multicolumn{1}{c|}{-2.845 } & \multicolumn{1}{c|}{99.42\%} & \multicolumn{1}{c|}{-2.769 } & \multicolumn{1}{c|}{94.24\%} & \multicolumn{1}{c|}{-2.808$\pm$0.035 } & 96.68$\pm$2.29\%  \\
    \hline
    20    & -2.652  & 86.70\% & \multicolumn{1}{c|}{-2.846 } & 99.80\% & \multicolumn{1}{c|}{-2.829 } & \multicolumn{1}{c|}{97.90\%} & \multicolumn{1}{c|}{-2.817 } & \multicolumn{1}{c|}{97.74\%} & \multicolumn{1}{c|}{-2.847 } & \multicolumn{1}{c|}{99.62\%} & \multicolumn{1}{c|}{-2.762 } & \multicolumn{1}{c|}{93.01\%} & \multicolumn{1}{c|}{-2.772 } & \multicolumn{1}{c|}{95.81\%} & \multicolumn{1}{c|}{-2.832 } & \multicolumn{1}{c|}{98.13\%} & \multicolumn{1}{c|}{-2.841 } & \multicolumn{1}{c|}{98.56\%} & \multicolumn{1}{c|}{-2.844 } & \multicolumn{1}{c|}{99.08\%} & \multicolumn{1}{c|}{-2.837 } & \multicolumn{1}{c|}{99.54\%} & \multicolumn{1}{c|}{-2.781 } & \multicolumn{1}{c|}{95.72\%} & \multicolumn{1}{c|}{-2.816$\pm$0.030 } & 97.51$\pm$1.98\%  \\
    \hline
    21    & -2.667  & 87.74\% & \multicolumn{2}{c|}{\multirow{30}[60]{*}{}} & \multicolumn{1}{c|}{-2.836 } & \multicolumn{1}{c|}{98.95\%} & \multicolumn{1}{c|}{-2.839 } & \multicolumn{1}{c|}{98.07\%} & \multicolumn{1}{c|}{-2.855 } & \multicolumn{1}{c|}{99.67\%} & \multicolumn{1}{c|}{-2.776 } & \multicolumn{1}{c|}{93.78\%} & \multicolumn{1}{c|}{-2.807 } & \multicolumn{1}{c|}{96.18\%} & \multicolumn{1}{c|}{-2.842 } & \multicolumn{1}{c|}{97.94\%} & \multicolumn{1}{c|}{-2.840 } & \multicolumn{1}{c|}{99.09\%} & \multicolumn{1}{c|}{-2.828 } & \multicolumn{1}{c|}{99.23\%} & \multicolumn{1}{c|}{-2.842 } & \multicolumn{1}{c|}{99.48\%} & \multicolumn{1}{c|}{-2.780 } & \multicolumn{1}{c|}{96.51\%} & \multicolumn{1}{c|}{-2.825$\pm$0.026 } & 97.89$\pm$1.78\%  \\
  \cline{1-3}\cline{6-27}    22    & -2.700  & 88.72\% & \multicolumn{2}{c|}{} & \multicolumn{1}{c|}{-2.843 } & \multicolumn{1}{c|}{99.44\%} & \multicolumn{1}{c|}{-2.835 } & \multicolumn{1}{c|}{98.00\%} & \multicolumn{1}{c|}{-2.852 } & \multicolumn{1}{c|}{99.74\%} & \multicolumn{1}{c|}{-2.797 } & \multicolumn{1}{c|}{95.47\%} & \multicolumn{1}{c|}{-2.798 } & \multicolumn{1}{c|}{97.26\%} & \multicolumn{1}{c|}{-2.830 } & \multicolumn{1}{c|}{98.91\%} & \multicolumn{1}{c|}{-2.845 } & \multicolumn{1}{c|}{99.03\%} & \multicolumn{1}{c|}{-2.836 } & \multicolumn{1}{c|}{99.33\%} & \multicolumn{1}{c|}{-2.844 } & \multicolumn{1}{c|}{99.74\%} & \multicolumn{1}{c|}{-2.803 } & \multicolumn{1}{c|}{97.47\%} & \multicolumn{1}{c|}{-2.828$\pm$0.020 } & 98.44$\pm$1.31\%  \\
  \cline{1-3}\cline{6-27}    23    & -2.714  & 90.18\% & \multicolumn{2}{c|}{} & \multicolumn{1}{c|}{-2.842 } & \multicolumn{1}{c|}{99.12\%} & \multicolumn{1}{c|}{-2.838 } & \multicolumn{1}{c|}{98.46\%} & \multicolumn{1}{c|}{-2.852 } & \multicolumn{1}{c|}{99.87\%} & \multicolumn{1}{c|}{-2.805 } & \multicolumn{1}{c|}{96.50\%} & \multicolumn{1}{c|}{-2.819 } & \multicolumn{1}{c|}{96.82\%} & \multicolumn{1}{c|}{-2.853 } & \multicolumn{1}{c|}{98.94\%} & \multicolumn{1}{c|}{-2.848 } & \multicolumn{1}{c|}{98.92\%} & \multicolumn{1}{c|}{-2.837 } & \multicolumn{1}{c|}{99.65\%} & \multicolumn{1}{c|}{-2.851 } & \multicolumn{1}{c|}{99.79\%} & \multicolumn{1}{c|}{-2.815 } & \multicolumn{1}{c|}{97.61\%} & \multicolumn{1}{c|}{-2.836$\pm$0.016 } & 98.57$\pm$1.15\%  \\
  \cline{1-3}\cline{6-27}    24    & -2.732  & 91.99\% & \multicolumn{2}{c|}{} & \multicolumn{1}{c|}{-2.842 } & \multicolumn{1}{c|}{99.25\%} & \multicolumn{1}{c|}{-2.842 } & \multicolumn{1}{c|}{98.78\%} & \multicolumn{1}{c|}{-2.847 } & \multicolumn{1}{c|}{99.85\%} & \multicolumn{1}{c|}{-2.814 } & \multicolumn{1}{c|}{96.69\%} & \multicolumn{1}{c|}{-2.834 } & \multicolumn{1}{c|}{97.78\%} & \multicolumn{1}{c|}{-2.849 } & \multicolumn{1}{c|}{99.02\%} & \multicolumn{1}{c|}{-2.849 } & \multicolumn{1}{c|}{99.22\%} & \multicolumn{1}{c|}{-2.844 } & \multicolumn{1}{c|}{99.54\%} & \multicolumn{1}{c|}{-2.848 } & \multicolumn{1}{c|}{99.64\%} & \multicolumn{1}{c|}{-2.809 } & \multicolumn{1}{c|}{98.57\%} & \multicolumn{1}{c|}{-2.838$\pm$0.014 } & 98.83$\pm$0.91\%  \\
  \cline{1-3}\cline{6-27}    25    & -2.740  & 93.46\% & \multicolumn{2}{c|}{} & \multicolumn{1}{c|}{-2.850 } & \multicolumn{1}{c|}{99.53\%} & \multicolumn{1}{c|}{-2.846 } & \multicolumn{1}{c|}{98.97\%} & \multicolumn{1}{c|}{-2.844 } & \multicolumn{1}{c|}{99.79\%} & \multicolumn{1}{c|}{-2.831 } & \multicolumn{1}{c|}{96.63\%} & \multicolumn{1}{c|}{-2.830 } & \multicolumn{1}{c|}{97.91\%} & \multicolumn{1}{c|}{-2.832 } & \multicolumn{1}{c|}{99.33\%} & \multicolumn{1}{c|}{-2.849 } & \multicolumn{1}{c|}{99.52\%} & \multicolumn{1}{c|}{-2.842 } & \multicolumn{1}{c|}{99.66\%} & \multicolumn{1}{c|}{-2.845 } & \multicolumn{1}{c|}{99.60\%} & \multicolumn{1}{c|}{-2.824 } & \multicolumn{1}{c|}{98.29\%} & \multicolumn{1}{c|}{-2.839$\pm$0.009 } & 98.92$\pm$0.97\%  \\
  \cline{1-3}\cline{6-27}    26    & -2.767  & 93.39\% & \multicolumn{2}{c|}{} & \multicolumn{1}{c|}{-2.847 } & \multicolumn{1}{c|}{99.65\%} & \multicolumn{1}{c|}{-2.844 } & \multicolumn{1}{c|}{99.04\%} & \multicolumn{1}{c|}{-2.857 } & \multicolumn{1}{c|}{99.86\%} & \multicolumn{1}{c|}{-2.817 } & \multicolumn{1}{c|}{97.22\%} & \multicolumn{1}{c|}{-2.829 } & \multicolumn{1}{c|}{98.41\%} & \multicolumn{1}{c|}{-2.852 } & \multicolumn{1}{c|}{99.35\%} & \multicolumn{1}{c|}{-2.850 } & \multicolumn{1}{c|}{99.45\%} & \multicolumn{1}{c|}{-2.835 } & \multicolumn{1}{c|}{99.46\%} & \multicolumn{1}{c|}{-2.846 } & \multicolumn{1}{c|}{99.58\%} & \multicolumn{1}{c|}{-2.836 } & \multicolumn{1}{c|}{98.96\%} & \multicolumn{1}{c|}{-2.841$\pm$0.011 } & 99.10$\pm$0.74\%  \\
  \cline{1-3}\cline{6-27}    27    & -2.767  & 93.71\% & \multicolumn{2}{c|}{} & \multicolumn{1}{c|}{-2.853 } & \multicolumn{1}{c|}{99.62\%} & \multicolumn{1}{c|}{-2.846 } & \multicolumn{1}{c|}{99.16\%} & \multicolumn{1}{c|}{-2.860 } & \multicolumn{1}{c|}{99.71\%} & \multicolumn{1}{c|}{-2.818 } & \multicolumn{1}{c|}{97.74\%} & \multicolumn{1}{c|}{-2.837 } & \multicolumn{1}{c|}{98.91\%} & \multicolumn{1}{c|}{-2.856 } & \multicolumn{1}{c|}{99.30\%} & \multicolumn{1}{c|}{-2.855 } & \multicolumn{1}{c|}{99.39\%} & \multicolumn{1}{c|}{-2.851 } & \multicolumn{1}{c|}{99.78\%} & \multicolumn{1}{c|}{-2.843 } & \multicolumn{1}{c|}{99.60\%} & \multicolumn{1}{c|}{-2.829 } & \multicolumn{1}{c|}{99.09\%} & \multicolumn{1}{c|}{-2.845$\pm$0.013 } & 99.23$\pm$0.57\%  \\
  \cline{1-3}\cline{6-27}    28    & -2.784  & 94.01\% & \multicolumn{2}{c|}{} & \multicolumn{1}{c|}{-2.846 } & \multicolumn{1}{c|}{99.86\%} & \multicolumn{1}{c|}{-2.850 } & \multicolumn{1}{c|}{99.44\%} & \multicolumn{1}{c|}{-2.853 } & \multicolumn{1}{c|}{99.79\%} & \multicolumn{1}{c|}{-2.831 } & \multicolumn{1}{c|}{98.19\%} & \multicolumn{1}{c|}{-2.845 } & \multicolumn{1}{c|}{98.76\%} & \multicolumn{1}{c|}{-2.841 } & \multicolumn{1}{c|}{99.64\%} & \multicolumn{1}{c|}{-2.845 } & \multicolumn{1}{c|}{99.57\%} & \multicolumn{1}{c|}{-2.846 } & \multicolumn{1}{c|}{99.63\%} & \multicolumn{1}{c|}{-2.836 } & \multicolumn{1}{c|}{99.55\%} & \multicolumn{1}{c|}{-2.839 } & \multicolumn{1}{c|}{99.15\%} & \multicolumn{1}{c|}{-2.843$\pm$0.006 } & 99.36$\pm$0.50\%  \\
  \cline{1-3}\cline{6-27}    29    & -2.775  & 95.09\% & \multicolumn{2}{c|}{} & \multicolumn{1}{c|}{-2.847 } & \multicolumn{1}{c|}{99.82\%} & \multicolumn{1}{c|}{-2.851 } & \multicolumn{1}{c|}{99.61\%} & \multicolumn{1}{c|}{-2.847 } & \multicolumn{1}{c|}{99.95\%} & \multicolumn{1}{c|}{-2.828 } & \multicolumn{1}{c|}{98.30\%} & \multicolumn{1}{c|}{-2.840 } & \multicolumn{1}{c|}{99.38\%} & \multicolumn{1}{c|}{-2.848 } & \multicolumn{1}{c|}{99.62\%} & \multicolumn{1}{c|}{-2.852 } & \multicolumn{1}{c|}{99.53\%} & \multicolumn{1}{c|}{-2.846 } & \multicolumn{1}{c|}{99.80\%} & \multicolumn{1}{c|}{-2.840 } & \multicolumn{1}{c|}{99.74\%} & \multicolumn{1}{c|}{-2.833 } & \multicolumn{1}{c|}{99.32\%} & \multicolumn{1}{c|}{-2.843$\pm$0.007 } & 99.51$\pm$0.44\%  \\
  \cline{1-3}\cline{6-27}    30    & -2.799  & 95.92\% & \multicolumn{2}{c|}{} & \multicolumn{1}{c|}{-2.842 } & \multicolumn{1}{c|}{99.82\%} & \multicolumn{1}{c|}{-2.855 } & \multicolumn{1}{c|}{99.33\%} & \multicolumn{1}{c|}{-2.853 } & \multicolumn{1}{c|}{99.88\%} & \multicolumn{1}{c|}{-2.843 } & \multicolumn{1}{c|}{98.41\%} & \multicolumn{1}{c|}{-2.844 } & \multicolumn{1}{c|}{99.51\%} & \multicolumn{1}{c|}{-2.850 } & \multicolumn{1}{c|}{99.63\%} & \multicolumn{1}{c|}{-2.843 } & \multicolumn{1}{c|}{99.62\%} & \multicolumn{1}{c|}{-2.845 } & \multicolumn{1}{c|}{99.70\%} & \multicolumn{1}{c|}{-2.842 } & \multicolumn{1}{c|}{99.69\%} & \multicolumn{1}{c|}{-2.850 } & \multicolumn{1}{c|}{99.38\%} & \multicolumn{1}{c|}{-2.847$\pm$0.005 } & 99.50$\pm$0.40\%  \\
  \cline{1-3}\cline{6-27}    31    & -2.809  & 95.41\% & \multicolumn{2}{c|}{} & \multicolumn{1}{c|}{-2.836 } & \multicolumn{1}{c|}{99.65\%} & \multicolumn{1}{c|}{-2.843 } & \multicolumn{1}{c|}{99.46\%} & \multicolumn{1}{c|}{-2.852 } & \multicolumn{1}{c|}{99.67\%} & \multicolumn{1}{c|}{-2.835 } & \multicolumn{1}{c|}{98.93\%} & \multicolumn{1}{c|}{-2.845 } & \multicolumn{1}{c|}{99.62\%} & \multicolumn{1}{c|}{-2.853 } & \multicolumn{1}{c|}{99.67\%} & \multicolumn{1}{c|}{-2.856 } & \multicolumn{1}{c|}{99.63\%} & \multicolumn{1}{c|}{-2.841 } & \multicolumn{1}{c|}{99.62\%} & \multicolumn{1}{c|}{-2.837 } & \multicolumn{1}{c|}{99.53\%} & \multicolumn{1}{c|}{-2.852 } & \multicolumn{1}{c|}{99.77\%} & \multicolumn{1}{c|}{-2.845$\pm$0.007 } & 99.55$\pm$0.22\%  \\
  \cline{1-3}\cline{6-27}    32    & -2.813  & 96.97\% & \multicolumn{2}{c|}{} & \multicolumn{1}{c|}{-2.840 } & \multicolumn{1}{c|}{99.88\%} & \multicolumn{1}{c|}{-2.838 } & \multicolumn{1}{c|}{99.46\%} & \multicolumn{1}{c|}{-2.854 } & \multicolumn{1}{c|}{99.73\%} & \multicolumn{1}{c|}{-2.838 } & \multicolumn{1}{c|}{98.84\%} & \multicolumn{1}{c|}{-2.849 } & \multicolumn{1}{c|}{99.57\%} & \multicolumn{1}{c|}{-2.846 } & \multicolumn{1}{c|}{99.60\%} & \multicolumn{1}{c|}{-2.850 } & \multicolumn{1}{c|}{99.61\%} & \multicolumn{1}{c|}{-2.845 } & \multicolumn{1}{c|}{99.85\%} & \multicolumn{1}{c|}{-2.846 } & \multicolumn{1}{c|}{99.70\%} & \multicolumn{1}{c|}{-2.839 } & \multicolumn{1}{c|}{99.76\%} & \multicolumn{1}{c|}{-2.844$\pm$0.005 } & 99.60$\pm$0.28\%  \\
  \cline{1-3}\cline{6-27}    33    & -2.810  & 96.58\% & \multicolumn{2}{c|}{} & \multicolumn{1}{c|}{-2.846 } & \multicolumn{1}{c|}{99.90\%} & \multicolumn{1}{c|}{-2.850 } & \multicolumn{1}{c|}{99.50\%} & \multicolumn{1}{c|}{-2.857 } & \multicolumn{1}{c|}{99.97\%} & \multicolumn{1}{c|}{-2.838 } & \multicolumn{1}{c|}{99.07\%} & \multicolumn{1}{c|}{-2.844 } & \multicolumn{1}{c|}{99.55\%} & \multicolumn{1}{c|}{-2.851 } & \multicolumn{1}{c|}{99.84\%} & \multicolumn{1}{c|}{-2.856 } & \multicolumn{1}{c|}{99.49\%} & \multicolumn{1}{c|}{-2.853 } & \multicolumn{1}{c|}{99.98\%} & \multicolumn{1}{c|}{-2.845 } & \multicolumn{1}{c|}{99.82\%} & \multicolumn{1}{c|}{-2.850 } & \multicolumn{1}{c|}{99.73\%} & \multicolumn{1}{c|}{-2.849$\pm$0.006 } & 99.69$\pm$0.27\%  \\
  \cline{1-3}\cline{6-27}    34    & -2.820  & 97.50\% & \multicolumn{2}{c|}{} & \multicolumn{1}{c|}{-2.848 } & \multicolumn{1}{c|}{99.74\%} & \multicolumn{1}{c|}{-2.842 } & \multicolumn{1}{c|}{99.68\%} & \multicolumn{1}{c|}{-2.845 } & \multicolumn{1}{c|}{99.98\%} & \multicolumn{1}{c|}{-2.841 } & \multicolumn{1}{c|}{99.35\%} & \multicolumn{1}{c|}{-2.846 } & \multicolumn{1}{c|}{99.54\%} & \multicolumn{1}{c|}{-2.838 } & \multicolumn{1}{c|}{99.64\%} & \multicolumn{1}{c|}{-2.850 } & \multicolumn{1}{c|}{99.63\%} & \multicolumn{1}{c|}{-2.844 } & \multicolumn{1}{c|}{99.81\%} & \multicolumn{1}{c|}{-2.850 } & \multicolumn{1}{c|}{99.56\%} & \multicolumn{1}{c|}{-2.841 } & \multicolumn{1}{c|}{99.76\%} & \multicolumn{1}{c|}{-2.845$\pm$0.004 } & 99.67$\pm$0.16\%  \\
  \cline{1-3}\cline{6-27}    35    & -2.828  & 97.97\% & \multicolumn{2}{c|}{} & \multicolumn{1}{c|}{-2.849 } & \multicolumn{1}{c|}{99.75\%} & \multicolumn{1}{c|}{-2.842 } & \multicolumn{1}{c|}{99.38\%} & \multicolumn{1}{c|}{-2.849 } & \multicolumn{1}{c|}{99.95\%} & \multicolumn{1}{c|}{-2.844 } & \multicolumn{1}{c|}{99.33\%} & \multicolumn{1}{c|}{-2.846 } & \multicolumn{1}{c|}{99.64\%} & \multicolumn{1}{c|}{-2.849 } & \multicolumn{1}{c|}{99.84\%} & \multicolumn{1}{c|}{-2.854 } & \multicolumn{1}{c|}{99.61\%} & \multicolumn{1}{c|}{-2.839 } & \multicolumn{1}{c|}{99.80\%} & \multicolumn{1}{c|}{-2.855 } & \multicolumn{1}{c|}{99.69\%} & \multicolumn{1}{c|}{-2.854 } & \multicolumn{1}{c|}{99.73\%} & \multicolumn{1}{c|}{-2.848$\pm$0.005 } & 99.67$\pm$0.19\%  \\
  \cline{1-3}\cline{6-27}    36    & -2.826  & 97.96\% & \multicolumn{2}{c|}{} & \multicolumn{22}{c|}{\multirow{15}[30]{*}{}} \\
  \cline{1-3}    37    & -2.833  & 98.23\% & \multicolumn{2}{c|}{} & \multicolumn{22}{c|}{} \\
  \cline{1-3}    38    & -2.829  & 98.27\% & \multicolumn{2}{c|}{} & \multicolumn{22}{c|}{} \\
  \cline{1-3}    39    & -2.835  & 98.37\% & \multicolumn{2}{c|}{} & \multicolumn{22}{c|}{} \\
  \cline{1-3}    40    & -2.839  & 98.62\% & \multicolumn{2}{c|}{} & \multicolumn{22}{c|}{} \\
  \cline{1-3}    41    & -2.847  & 98.84\% & \multicolumn{2}{c|}{} & \multicolumn{22}{c|}{} \\
  \cline{1-3}    42    & -2.833  & 98.87\% & \multicolumn{2}{c|}{} & \multicolumn{22}{c|}{} \\
  \cline{1-3}    43    & -2.841  & 99.07\% & \multicolumn{2}{c|}{} & \multicolumn{22}{c|}{} \\
  \cline{1-3}    44    & -2.841  & 99.13\% & \multicolumn{2}{c|}{} & \multicolumn{22}{c|}{} \\
  \cline{1-3}    45    & -2.843  & 99.25\% & \multicolumn{2}{c|}{} & \multicolumn{22}{c|}{} \\
  \cline{1-3}    46    & -2.844  & 99.12\% & \multicolumn{2}{c|}{} & \multicolumn{22}{c|}{} \\
  \cline{1-3}    47    & -2.846  & 99.26\% & \multicolumn{2}{c|}{} & \multicolumn{22}{c|}{} \\
  \cline{1-3}    48    & -2.855  & 99.12\% & \multicolumn{2}{c|}{} & \multicolumn{22}{c|}{} \\
  \cline{1-3}    49    & -2.843  & 99.43\% & \multicolumn{2}{c|}{} & \multicolumn{22}{c|}{} \\
  \cline{1-3}    50    & -2.844  & 99.62\% & \multicolumn{2}{c|}{} & \multicolumn{22}{c|}{} \\
    \hline
    \end{tabular}%
  }
  \label{tab:R90}%
\end{table}%

\clearpage
\begin{table}[hb]
  \centering
  \caption{Experimental results for the dissociation curve of the He–H$^+$ cation shown in Fig. 3. The energy, fidelity, and their errors are obtained by the average and standard deviation of results from ten repeated SPSA-QNG runs. In each run, the final energy is obtained using the average of measured energy values over five steps after convergence. After correcting for a constant systematic error, the data agree well with the theoretical energy, and the absolute error achieves chemical accuracy (1.5E-03 Hartree).} 
  \scalebox{0.9}{
    \begin{tabular}{|c|c|c|c|c|c|c|}
    \hline
    R     & Theoretical & \multicolumn{3}{c|}{SPSA-QNG} & \multicolumn{2}{c|}{Corrected} \\
    \cline{3-7}    ($\mathring{\text{A}}$)     & energy (MJ mol$^{-1}$) & Energy (MJ mol$^{-1}$) & Fidelity & Energy error (Hartree) & Energy (MJ mol$^{-1}$) & Energy error (Hartree)\\
    \hline
    0.4   & -2.372  & -2.362$\pm$0.001  & $99.89\pm0.02\%$  & 3.767E-03 & -2.375  & 1.185E-03 \\
    0.5   & -2.641  & -2.624$\pm$0.003  & $99.77\pm0.05\%$  & 6.433E-03 & -2.637  & 1.482E-03 \\
    0.7   & -2.831  & -2.817$\pm$0.002  & $99.74\pm0.06\%$  & 5.111E-03 & -2.830  & 1.600E-04 \\
    0.9   & -2.863  & -2.846$\pm$0.004  & $99.64\pm0.20\%$  & 6.262E-03 & -2.859  & 1.310E-03 \\
    1.1   & -2.854  & -2.841$\pm$0.002  & $99.65\pm0.12\%$  & 4.917E-03 & -2.854  & 3.428E-05 \\
    1.5   & -2.825  & -2.815$\pm$0.001  & $99.70\pm0.30\%$  & 3.656E-03 & -2.828  & 1.295E-03 \\
    2     & -2.811  & -2.801$\pm$0.009  & $99.12\pm1.01\%$  & 3.908E-03 & -2.814  & 1.043E-03 \\
    2.5   & -2.808  & -2.799$\pm$0.004  & $99.74\pm0.54\%$  & 3.627E-03 & -2.812  & 1.324E-03 \\
    3     & -2.808  & -2.799$\pm$0.009  & $98.60\pm1.35\%$  & 3.529E-03 & -2.809  & 2.799E-04 \\
    \hline
    \end{tabular}%
  }
  \label{tab:summary}%
\end{table}%

\section{Supplementary Figures}
\begin{figure}[h]
  \centering\includegraphics{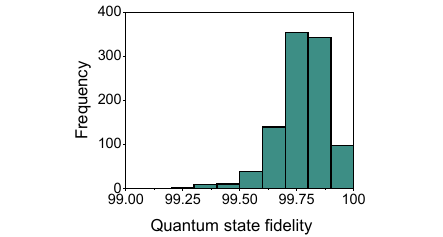}
  \caption{Histogram of the statistic quantum state fidelities for 1,000 randomly generated quantum states in the four-dimensional Hilbert space.}
  \label{fig: Fidelity}
\end{figure}

\begin{figure}[h]
  \centering\includegraphics{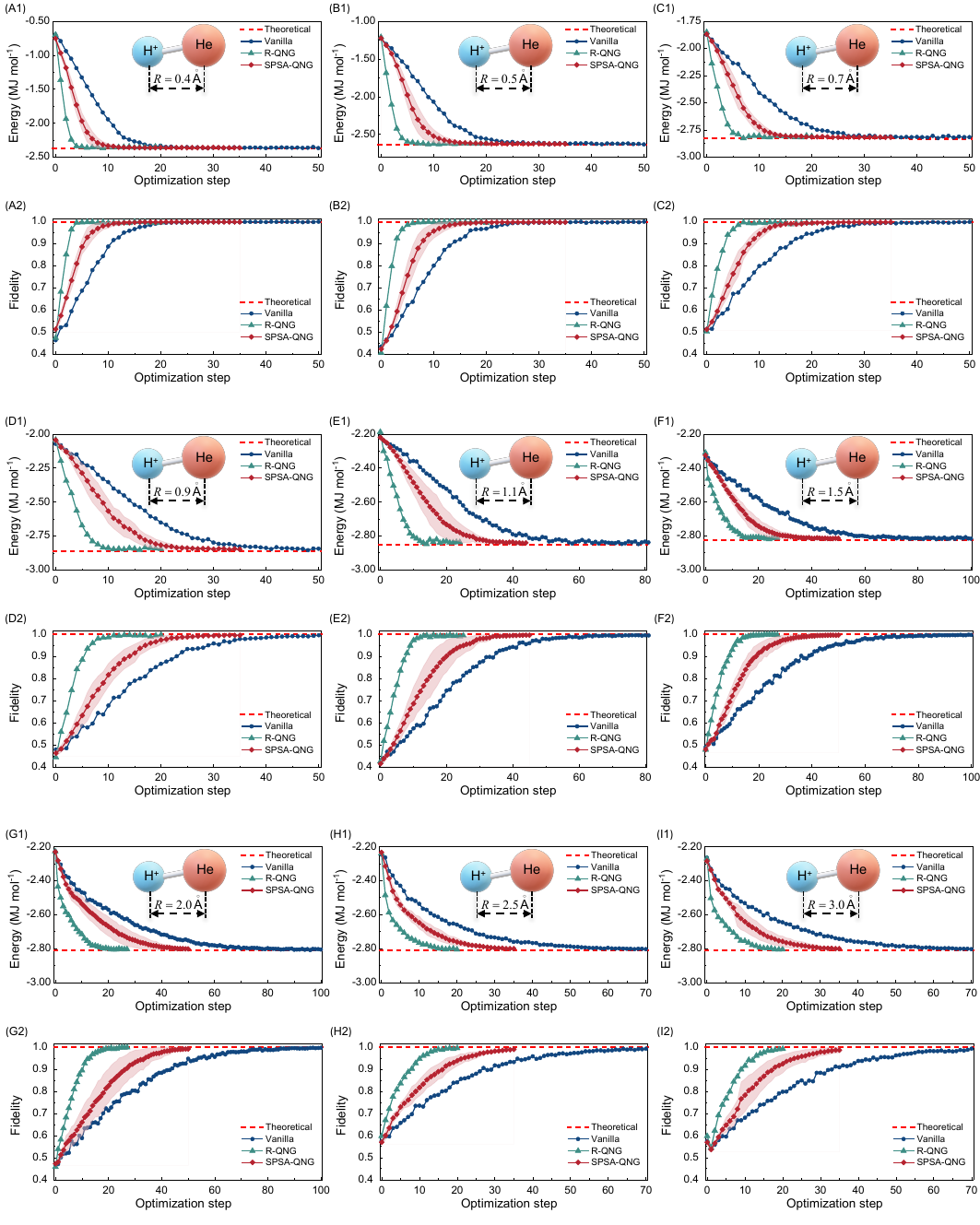}
  \caption{Finding the ground states of He–H$^+$ cation at the interatomic distance (A) $R=0.4 \, \mathring{\text{A}}$, (B) $R=0.5 \, \mathring{\text{A}}$, (C) $R=0.7 \, \mathring{\text{A}}$, (D) $R=0.9 \, \mathring{\text{A}}$, (E) $R=1.1 \, \mathring{\text{A}}$, (F) $R=1.5 \, \mathring{\text{A}}$, (G) $R=2.0 \, \mathring{\text{A}}$, (H) $R=2.5 \, \mathring{\text{A}}$, and (I) $R=3.0 \, \mathring{\text{A}}$. A comparison of the convergence performance of the evolution of  energy and fidelity with vanilla gradient descent (Vanilla), rigorous quantum natural gradient descent (R-QNG), and simultaneous perturbation stochastic approximated quantum natural gradient descent (SPSA-QNG) is presented. For SPSA-QNG, we repeat the optimization ten times, and the average and standard deviation (shaded area) of the energy and fidelity in the ten optimizations are plotted.}
  \label{fig: all R}
\end{figure}

\clearpage
